\documentclass[11pt,letter]{article}

\usepackage{amsmath}
\usepackage{xspace}
\usepackage{amssymb}
\usepackage{epsfig}

\newtheorem{Thm}{Theorem}
\newtheorem{Lem}[Thm]{Lemma}
\newtheorem{Cor}[Thm]{Corollary}

\newenvironment{proof}{\noindent {\textbf{Proof }}}{$\Box$ \medskip}

\newcommand\B{\{0,1\}}
\newcommand\Bn{\{0,1\}^n}

\newcommand {\ie} {\textit{i.e.}\xspace}
\newcommand {\st} {\textit{s.t.}\xspace}

\newcommand {\WP} {with probability\xspace}

\newcommand\ham[2]{\mbox{$\mathsf{Ham}_{#1}^{(#2)}$}\xspace}

\newcommand\parity{\mbox{$\mathsf{Parity}$}\xspace}
\newcommand\pr{\mbox{\bf Pr}}

\newcommand\alice{\mbox{\sf Alice}\xspace}
\newcommand\bob{\mbox{\sf Bob}\xspace}
\newcommand\referee{\mbox{\sf Referee}\xspace}

\begin{document}
\title{Tight bounds on the randomized communication complexity of symmetric XOR functions in one-way and SMP models\thanks{This work is supported by Hong Kong General Research Fund No. 419309 and No. 418710.}}
\author{Ming Lam Leung\thanks{Department of Computer Science and Engineering, The Chinese University of Hong Kong. Email: mlleung@cse.cuhk.edu.hk} \qquad Yang Li\thanks{Email: danielliy@gmail.com} \qquad Shengyu Zhang\thanks{Department of Computer Science and Engineering, The Chinese University of Hong Kong. Email: syzhang@cse.cuhk.edu.hk}}

\date{}
\maketitle

\abstract{
We study the communication complexity of symmetric XOR functions, namely functions $f: \{0,1\}^n \times \{0,1\}^n \rightarrow \{0,1\}$ that can be formulated as $f(x,y)=D(|x\oplus y|)$ for some predicate $D: \{0,1,...,n\} \rightarrow \{0,1\}$, where $|x\oplus y|$ is the Hamming weight of the bitwise XOR of $x$ and $y$. We give a public-coin randomized protocol in the Simultaneous Message Passing (SMP) model, with the communication cost matching the known lower bound for the \emph{quantum} and \emph{two-way} model up to a logarithm factor. As a corollary, this closes a quadratic gap between quantum lower bound and randomized upper bound for the one-way model, answering an open question raised in Shi and Zhang \cite{SZ09}.
}

\section{Introduction}
Communication complexity quantifies the minimum amount of communication needed for two (or sometimes more) parties to jointly compute some function $f$. Since introduced by Yao \cite{Yao79}, it has attracted significant attention in the last three decades, not only for its elegant mathematical structure but also for its numerous applications in other computational models \cite{KN97,LS09}.

The two parties involved in the computation, usually called \alice and \bob, can communicate in different manners, and here we consider the three well-studied models, namely the two-way model, the one-way model and the simultaneous message passing (SMP) model. In the two-way model, \alice and \bob are allowed to communicate interactively in both directions, while in the one-way model, \alice can send message to \bob and \bob does not give feedback to \alice. An even weaker communication model is the SMP model, where \alice and \bob are prohibited to exchange information directly, but instead they each send a message to a third party Referee, who then announces a result. A randomized protocol is called private-coin if \alice and \bob each flip their own and private random coins. If they share the same random coins, then the protocol is called public-coin. The private coin model differs from the public coin model by at most an additive factor of $O(\log n)$ in the two-way and one-way models \cite{New91}. 

We use $R^{priv}(f)$ to denote the communication complexity of a best private-coin randomized protocol that computes $f$ with error at most $1/3$ in the two-way protocol. Similarly, we use the $R^{||,priv}(f)$ to denote the communication complexity in the private-coin SMP model, and $R^{1,priv}(f)$ for the private-coin one-way model. Changing the superscript ``priv" to ``pub" gives the notation for the communication complexities in the public-coin models. If we allow \alice and \bob to use quantum protocols, then $Q(f), Q^{1}(f), Q^{||}(f)$ represent the quantum communication complexity in two-way model, one-way model and SMP model, separately. In the quantum case the communication complexity is evaluated in terms of the number of qubits in the communication. If \alice and \bob share prior entanglement, then we use a star in the superscript to denote the communication complexity.

Arguably the most fundamental issue in communication complexity is to determine the largest gap between the quantum and classical complexity. In particular, there is no super-constant separation between quantum and classical complexities in the one-way model; actually, it could well be the truth that they are the same up to a constant factor for all total Boolean functions. 

One way to understand the question is to study special classes of functions. An important class of Boolean functions is that of XOR functions, namely those in the form of $f(x\oplus y)$ where $x\oplus y$ is the bitwise XOR of $x$ and $y$. Some well studied functions such as the Equality function and the Hamming Distance function are special cases of XOR functions. XOR functions belong to a larger class of ``composed functions"; see \cite{LZ10} for some recent studies.

While the general XOR function seems hard to study, recently Shi and Zhang \cite{SZ09} considered symmetric XOR functions, \ie  $f(x\oplus y)=D(|x\oplus y|)$ for some $D: \{0,1,...,n\} \rightarrow \{0,1\}$. Define $r_0$ and $r_1$ to be the minimum integers such that $r_0,r_1\le n/2$ and $D(k)=D(k+2)$ for all $k\in [r_0,n-r_1)$ and set $r=\max \{r_0,r_1\}$. Shi and Zhang proved that the quantum lower bound for symmetric XOR functions in the two-way model is $\Omega(r)$, and on the other hand, they also gave randomized protocol in communication of $\tilde{O}(r)$ in the two-way model and $\tilde{O}(r^2)$ in the one-way model. Pinning down the quantum and randomized communication complexity of symmetric XOR functions in the one-way model was raised as an open problem.

In this work, we close the quadratic gap by proving a randomized upper bound of $\tilde O(r)$, which holds even for the SMP model. Namely,

\begin{Thm}\label{thm: rub}
For any symmetric XOR function $f$,
\begin{equation}
	R^{\|,pub}(f)=O(r{\log^3 r}/\log{\log r})
\end{equation}
\end{Thm}

Combining this upper bound with Shi and Zhang's quantum lower bound in the two-way model, we have the following.
\begin{Cor}
The randomized and quantum communication complexities of symmetric XOR functions are $\tilde{\Theta}(r)$, in the two-way, the one-way and the public-coin SMP models.
\end{Cor}

A good question for further exploration is the private-coin SMP model.

\section{Preliminaries}
In this part we review some known results on the randomized and quantum communication complexity of the Hamming Distance function and the Equality function. 

Let $\ham{n}{d}$ be the boolean function such that $\ham{n}{d}(x,y)=1$ if and only if the two $n$-bit strings $x$ and $y$ have Hamming distance at most $d$. Yao \cite{Yao03} showed a randomized upper bound of $O(d^2)$ in the public-coin SMP model, lated improved by Gavinsky, Kempe and de Wolf \cite{GKW04} to $O(d\log n)$ and further by Huang, Shi, Zhang and Zhu \cite{HSZZ06} to $O(d\log d)$. Let $\mathcal{HD}_{d,\epsilon}$ denote the $O(d\log d \log(1/\epsilon))$-cost randomized protocol by repeating the \cite{HSZZ06} protocol for $O(\log(1/\epsilon))$ times so that the error probability is below $\epsilon$. 


The parity function $\parity(x)$ is defined as $\parity(x)=1$ if and only if $|x|$ is odd.

A function $f:\Bn\times \Bn \rightarrow \B$ is a symmetric XOR function if $f(x,y) = S(x\oplus y)$ for some symmetric function $S$. That is, $f(x,y) = D(|x\oplus y|)$ where $D:\{0, 1, \ldots, n\}\rightarrow \B$. Let $\tilde D(k) = D(n-k)$ and $\tilde S(x,y) = \tilde D(|x\oplus y|)$. Define $r_0$ and $r_1$ to be the minimum integers such that $r_0,r_1\le n/2$ and $D(k)=D(k+2)$ for all $k\in [r_0,n-r_1)$; set $r=\max \{r_0,r_1\}$. By definition, $D(k)$ only depends on the parity of $k$ when $k\in [r_0, n-r_1]$. Suppose $D(k) = T(\parity(k))$ for $k\in [r_0, n-r_1]$ (for some function $T$).

All the logarithms in this paper are based 2. 
\section{A public-coin protocol in the SMP model}
This section gives the protocol in Theorem \ref{thm: rub}. We will first give a subprocedure $\mathcal P_k$ which computes the function in the special case of $|x\oplus y| \leq k$. It is then used as a building block for the general protocol $\mathcal P$. 

In the protocols we will use random partitions. A \emph{random $k$-partition} of $[n]$ is a random function $p$ mapping $[n]$ to $[k]$, \ie mapping each element in $[n]$ to $[k]$ uniformly at random and independently. We call the set $\{i\in [n]: p(i) = j\}$ the \emph{block} $B(j)$. A simple fact about the random partition is the following. 

\begin{Lem}\label{lem: rand part}
	For any string $z\in \B^n$ with at most $k$ 1's, a random $k$-partition has 
	\begin{equation}
		\pr[\text{All $k$ blocks have less than $c$ 1's}] \geq 1 - O(1/k^2).
	\end{equation} 
	where $c = 4 \log k / \log \log k$.
\end{Lem}

\begin{proof}
	Consider the complement event. There are $k$ possible blocks to violate the condition, $\binom{k}{c}$ choices for the $c$ 1's (out of $k$ 1's) put in the ``bad" block, and for each of these 1's, the probability of it mapped to the block is $1/k$. Thus the union bound gives
	\begin{align}
		 \pr[\text{There exists a block with $c$ 1's}]
		\leq  k \cdot \binom{k}{c} \cdot \frac{1}{k^c} 
		\leq  \big(\frac{ek}{c} \big)^c  \cdot \frac{1}{k^{c-1}} = \big(\frac{e}{c} \big)^c  \cdot k 
	\end{align}
	It is easily verified that the chosen $c$ makes this bound $O(1/k^2)$.
\end{proof}

Now the protocol $P_k$ is as in {\bf  Box $\mathcal P_k$}. Recall that $\mathcal{HD}_{d,\epsilon}$ is the $O(d\log d \log(1/\epsilon))$ randomized protocol with error probability below $\epsilon$.

\begin{center}
\fbox{
\begin{minipage}[l1pt]{5in}
{\bf  Box $\mathcal P_k$:}\\

{\bf  A public-coin randomized protocol $\mathcal P_k$ for functions $f(x,y) = D(|x\oplus y|)$, with promise $|x\oplus y| \leq k$, in the SMP model} \\

{\bf Input}: $x\in \Bn$ to \alice and $y\in \Bn$ to \bob, with promise $|x\oplus y| \leq k$ \\
{\bf Output}: One bit $\bar f$ by \referee satisfying $\bar f = f(x,y)$ with probability at least 0.9. \\

\vspace{1em}
{\bf Protocol}:

\alice and \bob: 
\begin{enumerate}
	\item Use public coins to generate a common random $k$-partition $[n] = \uplus_{i=1}^k B(i)$.
	\item {\bf for} $i$ = 1 \textbf{to} $k$ 
	
	\quad {\bf for} $j$ = 0 \textbf{to} $c = 4 \log k / \log \log k$ 
	
	\quad \quad run (\alice and \bob's part of) the protocol $\mathcal{HD}_{j,\epsilon}$ on input $(x_{B(i)}, y_{B(i)})$ with $\epsilon = 1/(10k \log c)$, sending a pair of messages $(m_{a,i,j}(x_{B(i)})$, $m_{b,i,j}(y_{B(i)}))$. 
\end{enumerate}

\vspace{.5em}
\referee:
\begin{enumerate}
	\item {\bf for} $i$ = 1 \textbf{to} $k$ 
	
	\begin{enumerate}
		\item On receiving $\{(m_{a,i,j}(x_{B(i)}), m_{b,i,j}(y_{B(i)})): j = 1, \ldots, c\}$, run (\referee's part of) the protocol $\mathcal{HD}_{j,\epsilon}$ which outputs $h_{ij}$.
		\item Use binary search in $(h_{i1}, \ldots, h_{ic})$ to find the Hamming distance $h_i$ of $(x_{B(i)}, y_{B(i)})$.
	\end{enumerate}
	\item Output $D(\sum_{i=1}^k h_i)$.
\end{enumerate}
\end{minipage}
}
\end{center}

\begin{Lem}
	If $|x \oplus y| \leq k$, then \referee outputs $D(|x\oplus y|)$ \WP at least 0.9. The cost of protocol $\mathcal{P}_k$ is $O(k\log^3 k/\log\log k)$.
\end{Lem}
\begin{proof}
	First, by Lemma \ref{lem: rand part}, each block contains at most $c$ different indices $i$ \st $x_i \neq y_i$. Namely, the Hamming distance of $x_{B(i)}$ and $y_{B(i)}$ is at most $c$. Thus running the protocols $\mathcal{HD}_{j,\epsilon}$ for $j = 0, ..., c$ would give information to find the Hamming distance $h_i$ of $(x_{B(i)}, y_{B(i)})$. In each block $B(i)$, $h_i$ is correctly computed as long as each of the $\lceil \log c \rceil$ values $h_{ij}$ on the (correct) path of the binary search is correct. Thus a union bound gives the overall error probability upper bounded by $k (\log c) \epsilon = 1/10$. The cost of the protocol is $O(k\cdot c\cdot c\log c\log (1/\epsilon)) = O(k\log^3 k/\log\log k)$.
\end{proof}

With the protocol $\mathcal{P}_k$ in hand, we now construct the general protocol as in {\bf Box $\mathcal{P}$.}

\begin{center}
\fbox{
\begin{minipage}[l1pt]{5.00in}
{\bf Box $\mathcal P$}\\

{\bf  A public-coin randomized protocol $\mathcal P$ for functions $f(x,y) = S(x\oplus y)$ in the SMP model} \\

{\bf Input}: $x\in \Bn$ to \alice and $y\in \Bn$ to \bob \\
{\bf Output}: One bit $b$ which equals to $f(x,y)$ with probability at least 2/3.

\vspace{1em}
{\bf Protocol}:

\begin{enumerate}
	\item Run the protocol $\mathcal{HD}_{r_0, 1/10}$ and the protocol $\mathcal{HD}_{n-r_1, 1/10}$ on $(\bar x,y)$.
	\item Run the protocol $\mathcal{P}_{r_0}$ for function $S$ on $(x,y)$ and the protocol $\mathcal{P}_{r_1}$ for function $\tilde S$ on $(\bar{x},y)$.
	\item \alice: send $\parity(x)$ 
	\item \bob: send $\parity(y)$.
	\item \referee: 
	\begin{enumerate}
		\item If $\mathcal{HD}_{r_0, 1/10}$ on $(x,y)$ outputs 1, then output $\mathcal{P}_{r_0}$ on $(x,y)$ and halt.
		\item If $\mathcal{HD}_{n-r_1, 1/10}$ on $(\bar{x},y)$ outputs 1, then output $\mathcal{P}_{r_1}$ on $(\bar{x},y)$ and halt. 
		\item Output $T(\parity(x)\oplus \parity(y))$.
	\end{enumerate} 
\end{enumerate}
\end{minipage}
}
\end{center}

\begin{Thm}
	The protocol $\mathcal{P}$ outputs the correct value with probability at least $2/3$, and the complexity cost is $O(r\log^3 r/\log\log r)$. 
\end{Thm}
\begin{proof}
\emph{Correctness}: 
If $|x\oplus y| \leq r_0$, then \WP at least 0.9, the protocol $\mathcal{HD}_{r_0,1/10}(x,y)$ outputs 1, thus \referee outputs $\mathcal{P}_{r_0}$ on $(x,y)$, which equals to $f(x,y)$ \WP at least 0.9 by the correctness of the protocol $\mathcal{P}_{r_0}$. Thus the overall success probability is at least $0.81 > 2/3$. 
	
If $|x\oplus y| \geq n-r_1$, then $|\bar{x}\oplus y| \leq r_1$ and \WP at least 0.9, the protocol $\mathcal{HD}_{r_1,1/10}(\bar{x},y)$ outputs 1, thus \referee outputs $\mathcal{P}_{r_1}(\tilde S, \bar{x},y)$, which equals to 
\begin{equation}
	 \tilde S(\bar x \oplus y) = \tilde D(n- |x \oplus y|) = D(|x \oplus y|)
\end{equation}
\WP at least 0.9 by the correctness of the protocol $\mathcal{P}_{r_1}$. Thus the overall success probability is at least $0.81 > 2/3$. 
	
	If $r_0 < |x\oplus y| < n-r_1$, then the protocol proceeds to the very last step with probability at least $1-0.1-0.1 = 0.8$. And once this happens, then \referee outputs the correct value with certainty, since $f(x,y) = T(\parity(x\oplus y)) = T(\parity(x) \oplus \parity(y))$. 
	
	\emph{Complexity}: The cost is twice of the cost of the protocol $P_r$, plus twice of the cost of the protocol $\mathcal{HD}_{r,1/10}$, plus 2, which in total is $O(r\log^3 r/\log\log r)$.
\end{proof}

\bibliography{XORSMP}
\bibliographystyle{plain}
\end{document}